\documentclass[12pt]{article}
\usepackage[dvips]{graphicx}

\input tcilatex2.5

\QQQ{Language}{
American English
}

\begin{document}


\title{Expansion and evaporation of hot nuclei: Comparison between 
semi-classical
and quantal mean-field approaches }
\author{D.~Lacroix$^1$, Ph.~Chomaz$^1$ \\
\medskip (1)G.A.N.I.L., B.P. 5027, F-14076 Caen Cedex 5, France.}
\maketitle

\begin{abstract}
We present a general discussion of the mean field dynamics of finite nuclei
prepared under extreme conditions of temperature and pressure 
We compare the prediction of semi-classical approximation with complete
quantum simulations. Many features of the dynamics are carefully studied
such as the collective expansion, the evaporation process, the different
time-scale... This study points out many quantitative differences between
quantum and semi-classical approaches. Part of the differences are
related to numerical features inherent in semi-classical simulations but
most of them are a direct consequence of the non treatment of nuclei as
quantal objects. In particular, we show that because of a too strong damping
in semi-classical approaches the expansion of hot nuclei is quenched and the
speed of the collective motion reduced.
\end{abstract}

\smallskip

 
\smallskip


\newpage

\section{\protect\medskip Introduction}

Semi-classical methods are widely used to describe the nuclear dynamics. In
particular, considerable progresses have been made in semi-classical transport
theory of Heavy Ion Collisions with the simulation of Boltzmann type, in
which an averaged Pauli-blocked collision term is added on top of a
mean-field evolution\footnote{{\footnotesize These approaches are often
referred under different acronyms BUU, VUU, BNV\cite{Gal85,Gre87,Bau87,Toh92}... In this article we will
use BUU (for Boltzmann-Ueling-Ulenbeck) which is the one initially proposed
by Bertsch et al \cite{ber1,ber2,Bon}}} \cite{ber1,ber2,Bon}. More recently,
following the Boltzmann-Langevin (BL) theory, a stochastic term has been
also introduced to take into account the strong fluctuations of observables 
\cite{fluc1}. These approaches (BUU and BL) are of great interest not only
because they are numerically tractable but also because they seem to be
globally in good agreement with experimental data.

As far as quantum features are concerned, semi-classical approaches
essentially take into account the exclusion principle through the Pauli
blocking in nucleon-nucleon collisions. Thus, many properties of the nucleus
which are due to the quantum nature of the nuclear systems\cite{BM}, might
not be correctly described in semi-classical approximations.
For example, delocalization of wave-packets, wave dynamics, barrier
transmission... are absent from semi-classical approaches, it is thus of
great interest to check if the loss of these quantal features of nucleons
does not change the evolution of a highly excited nucleus. The validity
of semi-classical theories was already discussed several times and some
qualitative differences have already been pointed-out\cite{Tang}. However,
we want to present a systematic quantitative comparison for 
the desexcitation of nuclei under extreme conditions.

In this paper, we compare the evolution of a hot and compressed (or diluted)
nucleus in quantal calculation with the results of a semi-classical
simulation. Our aim is to bring quantitative information about the possible
role of the quantal nature of the nucleons and about the accuracy of
semi-classical approaches. Since, most of the discussed phenomena are
already present at the mean-field level, we will restrict ourselves to a
comparison between time dependent Hartree-Fock (TDHF) methods and its
semi-classical counterpart: the Vlasov dynamics. Moreover, we will consider
a simple scenario of the evolution of a hot spherical compound nucleus and
we will analyze evaporation of particles, dynamics of collective motion,
time-scales...

In section 2, we present the considered scenario together with the various
ingredients entering in the mean-field approximation.
We will see that some of the differences between quantum and semi-classical
simulations could be related to the finite sampling of the phase-space
density, necessary in semi-classical methods. This numerical sampling of the
phase-space generates a spurious finite range term in the semi-classical
mean-field which was not in the original quantal description. This term
modifies the properties of a finite piece of nuclear matter. When the
quantum potential has a finite range component one may try to reduce the
spurious effects by artificially reducing the range of the interaction used
in the semi-classical simulations. We will show that this method is hardly a
good approximation especially due to the density dependence of the various
term in the interaction. In order to disentangle the role of this numerical
artifact from a real difference between a quantal and a semi-classical
treatment, a third type of simulation will be displayed: we will consider a
TDHF simulation where a surface term has been added to the interaction in
order to mimic the semi-classical numerical range.

In section 3, we present results concerning the expansion dynamics 
of initial excited nuclei with 
different compression or dilatation. 

\smallskip


As expected, the semi-classical
simulation seems to be a better approximation of the TDHF simulation where
the surface term has been introduced. Nevertheless, many differences remain,
showing limits of semi-classical calculations. For instance, evaporation
rate, dynamics of the collective motion and associated time-scales are
different. The condition under which a quantum system will expand toward low
density regions will be finally discussed.

\section{A scenario to compare mean-field approaches.}

A good benchmark for a comparison between semi-classical and quantum
dynamics is provided by the model case introduced by D. Vautherin et al\cite
{Vau1}. Indeed, they consider an excited spherical nucleus at various
initial temperature and with various density profiles. Then, they follow the
evolution in the TDHF framework. The argument in favor of the use of a
mean-field evolution is that the nucleon-nucleon collisions
are not important for the dynamics of such thermalized system.
Indeed, in the case of an equilibrated expanding source the gain and
the loss terms in the collision part of the BUU approach almost cancel.


\subsection{\protect\smallskip The interaction}

We will study the self-consistent mean field evolution of a
spherical nucleus under extreme conditions. In the first part of this
article we shall consider a density dependent effective two-body
interaction: 
\begin{equation}
\hat{V}(1,2)=t_0~\delta (\hat{\vec{r}_1}-\hat{\vec{r}_2})+t_3~\rho \left( 
\frac{\hat{\vec{r}_1}+\hat{\vec{r}_2}}2\right) ^\sigma \delta (\hat{\vec{r}_1%
}-\hat{\vec{r}_2}),  \label{eq:inter}
\end{equation}

\noindent which is a simplified version of the more general Skyrme force,
but has been shown to be sufficient for the study of the expansion dynamics%
\cite{Vau1}. In the mean-field approach, for spin-isospin saturated systems,
this interaction corresponds to a local potential energy 
\begin{equation}
E\left[ \hat{\rho}\right] (r,t)=\frac 38t_0\rho ^2(r,t)+\frac 1{16}\rho
^{2+\sigma }(r,t),  \label{eq:E}
\end{equation}
Note that, in the second part of this article, we shall add more terms to
this interaction in order to take into account the finite range of the
nuclear forces. Many of the useful results and derivations considering
mean-field for a Skyrme force are derived in ref.\cite{Vau2}.

For our numerical applications, we have taken $t_0=-1000MeVfm^3$, $%
t_3=15000MeVfm^6$ and $\sigma =1$ (see ref.\cite{Vau1}). 
In the infinite
nuclear matter, this potential is globally in agreement with some more
complicated parametrization\cite{Blaizot}: this particular force yields to
the following saturation point properties: $E/A=-14MeV$, a saturation
density $\rho _0=0.15fm^{-3}$ and a compression modulus $K_\infty =336.2MeV$.
This value corresponds to a rather stiff equation of state. Another
interesting quantity is the ''zero temperature spinodal density'' , 
this is the
density where the compressibility drops to zero at zero temperature. For the
considered force, this density is equal to $0.1fm^{-3}$.

\smallskip

\subsection{Constrained mean-field initialization}

\smallskip In simulation of mean field type, the usual way for preparing
systems at different temperatures and with different shapes $\prec \hat{Q}%
\succ $, where $\hat{Q}$ is the observable measuring the deformation, 
is to perform a constrained mean-field calculation with a constraining
field $\lambda \hat{Q}$. 
For instance, $\hat{Q}$ can be the
square radius operator 
and
$\lambda $ is computed in order to obtain the desired square radius $\prec 
\hat{r}^2\succ $ or equivalently the desired compression factor $\eta
=\left( \prec \hat{r}^2\succ _0/\prec \hat{r}^2\succ \right) ^{3/2}$ (where $%
\prec \hat{r}^2\succ _0$ is the RMS of the ground state at zero
temperature). We will also define the compression using the average density (%
$\rho $) in a sphere of radius 2fm around the center of the nucleus. 
The value of 2 fm is small enough
compared to the RMS to avoid the bias of the surface and big enough 
to take into account not only the central 
s-orbitals but also others in order to
focus on collective effects.

\smallskip 
The temperature can be either fixed arbitrary to study isothermal initial
conditions or can be defined in order to get a given energy $\prec \hat{H}%
\succ $. This method is well adapted to generate isothermal initial
conditions however when the nucleus is strongly compressed, the
self-consistency of the mean-field may lead to some shape modification such
as the apparition of hollow structures. Moreover, it appears rather
difficult to generate diluted nuclei since negative $\lambda $ values leads
to unbound systems.

\subsection{Isoentropic scaling initialization}

In order to overcome the limitations of the constrained mean-field approach
and to be able to reach both high or low densities we have considered an
alternative procedure based on the scaling of the nucleus. Since the scaling
preserves the entropy, it is more natural to consider isoentropic families
of initial nuclei.

\smallskip In order to define the initial density before scaling, we can
use the constrained mean field method described above with a small
constraint parameter $\lambda $ introduced to correctly treat the continuum (%
$\lambda =0.25MeVfm^{-2}$ in our simulation). The initial
temperatures $T$ are defined in order to get the desired entropies $S$.

For numerical reasons, we have chosen to
neglect the temperature dependence of the mean-field potential 
in the solution of
the constrained mean-field. Therefore, the first stage of the initialization
procedure is to find the static solution $\rho _0$ of the mean field problem
at zero temperature with a small external constraint $\lambda r^2$. This
provides us with a well defined one-body potential $U\left[ \rho _0\right] $%
. Then, we have constructed a finite-temperature particle distribution as a
hot gas of independent particles in this self-consistent potential, $U\left[
\rho_0\right] $. Then using the scaling 
\begin{eqnarray}
\vec{r}\longrightarrow \vec{r^{\prime }} &=&\alpha \vec{r},
\label{eq:scaling1} \\
\vec{p}\longrightarrow \vec{p^{\prime }} &=&\frac{\vec{p}}\alpha .
\label{eq:scaling2}
\end{eqnarray}
which conserves the entropy, we can generate various families of
isoentropic initial conditions with different dilutions or compressions.
%
%

\subsection{\protect\smallskip Alternative approximate isothermal
initialization}

If we do not take into account the self-consistent part of the mean-field
potential, we see that energies scale approximately as 
\begin{equation}
\varepsilon \longrightarrow \varepsilon ^{\prime }\simeq \varepsilon /\alpha
^2,  \label{eq:scaling4}
\end{equation}
showing that the temperatures approximately depend on $\alpha $ as 
\begin{equation}
T\longrightarrow T^{\prime }\simeq T/\alpha ^2,  \label{eq:T_scale}
\end{equation}

\smallskip Therefore we can also generate a family of approximately
isothermal initial densities by keeping $T^{\prime }$ constant, i.e.
changing the temperature $T$ prior to the scaling. Results obtained with
this approximation are in all points comparable to Constrained Hartree-Fock
(CHF) simulations for a large range of initial conditions within the
limitations of CHF discussed above. This gives good confidence
in the scaling method and in the robustness of the conclusions we draw.
Moreover, we would like to stress that, in fact, any initial conditions can
be studied since the most important point is to use identical protocols in
quantum and classical simulations so that the comparison is meaningful.

\section{\protect\smallskip \protect\smallskip Quantum model}

Following \cite{Vau1}, we consider spherically symmetric nuclei. The
one-body wave functions can be separated into their radial part, angular part
and spin-isospin part

\begin{equation}
\Phi _\alpha (\vec{r},s,t)=\frac{R_{nl}(r)}rY_{lm}(\theta ,\varphi )\chi
_{\sigma,\tau}(s,t)  \label{eq:Phi}
\end{equation}

\noindent in this expression $\vec{r}=(r,\theta ,\varphi )$, $\chi _{\sigma
\tau }$ represents the spin and isospin part. In our special case, we
consider that our system is spin and isospin saturated. Finally, $\alpha $
is an abbreviation for all the quantum number $(n,l,m,\sigma,\tau)$. 

For the considered spherical system, the TDHF equation reduces to a radial
equation for the single particle wave functions 
\begin{equation}
i\hbar \frac{\partial R_{nl}(r,t)}{\partial {t}}=\left\{ -\frac{\hbar ^2}{2m}%
\frac{\partial ^2}{\partial {r^2}}+\frac{\hbar^2 l(l+1)}{2mr^2}+U\left[ \rho
\right] (r,t)+V_{Coulomb}\right\} R_{nl}(r,t),  \label{eq:evol}
\end{equation}
In this equation $\hat{V}_{coulomb}$ is the Coulomb potential\footnote{In this 
calculation the coulomb field is considered in
an approximate way by giving the charge $+1/2$ to each nucleons.}
while $U$ is
the density-dependent mean-field potential 
\begin{equation}
\hat{U}\left[ {\rho}\right] (r,t)=\frac 34t_0\rho (r,t)+\frac{2+\sigma }{16}%
\rho ^{1+\sigma }(r,t),  \label{eq:Pot}
\end{equation}
In this expression, $\rho$ is the local density
\begin{equation}
\rho (r,t)=4\sum_{n,l}(2l+1)n_{nl}\frac{\left| R_{nl}(r,t)\right| ^2}{4\pi
r^2},  \label{eq:Rho2}
\end{equation}
where $n_{nl}$ are the time-independent occupation numbers which only depend 
upon $n$ and $l$ quantum numbers. The factor $4$ is coming from the spin and
isospin degeneracy. 

\subsection{\protect\smallskip Quantum initialization}

As discussed in the introduction, we have studied different types of initial
conditions consisting in varying the dilution, the temperature and the
entropy of the considered nuclei.

\subsubsection{\protect\smallskip Constrained Hartree-Fock}

In the constrained Hartree-Fock approach, the occupation numbers $n_{nl}$
are distributed according to the Fermi-Dirac distribution

\begin{equation}
n_{nl}(T)=\frac 1{\exp \left( \frac{\varepsilon _{nl}-\mu }T\right) +1}
\label{eq:fermi}
\end{equation}
where the energies $\varepsilon _{nl}$ are the eigenvalue of the constrained
Hamiltonian $h+\lambda r^2$ while the single particle wave functions are
nothing but the eigenstates of the constrained Hamiltonian. The parameters $%
\lambda $ and $\mu $ are defined requiring a given dilution and a given
number of particles.

\subsubsection{Scaling initialization}

In the second method, we use the energy levels and the wave functions of the
zero temperature mean field in presence of a small external field ($\lambda
r^2$, $\lambda =0.25MeVfm^{-2}$). This gives the energies $\varepsilon _{nl}$
and the associated single particle wave functions $R_{nl}$ . Then, these
states are occupied according to the Fermi-Dirac distribution (\ref{eq:fermi}%
). The temperature, $T$, is computed in order to get the desired entropy 
\begin{equation}
S=-4\sum_{n,l}(2l+1)\left( n_{nl}\log n_{nl}+(1-n_{nl})\log (1-n_{nl})\right)
\label{eq:S}
\end{equation}
Finally, in order to get the appropriate dilution, the wave functions are
rescaled by the transformation 
\begin{equation}
R_{nl}^{\prime }(r^{\prime })=\frac{R_{nl}(\frac{r^{\prime }}\alpha )}{%
\alpha ^{3/2}},  \label{eq:scaling5}
\end{equation}

This initialization procedure can also be used in order to generate scaled
density at approximately constant temperature by scaling the temperature $T$
according to Eq. (\ref{eq:T_scale}).

\subsection{\protect\smallskip Numerical details}

In practice, the self-consistent solution of the static Hartree Fock (at
finite temperature and with a constraint $\lambda r^2$) has been found
through an iterative application of the operator $1/\left( \varepsilon
_{nl}-\left( \hat{h}+\lambda \hat{r}^2\right) \right) $ to the single
particle states $\Phi _{nl}$. In this procedure, at each step the one-body
density is computed using the occupation numbers (\ref{eq:fermi}) where the
energy is given by $\varepsilon _{nl}=\left\langle \Phi _{nl}\right| \hat{h}%
+\lambda \hat{r}^2\left| \Phi _{nl}\right\rangle $. We start with the
oscillator wave functions and after many iterations, we converge to the
solution of our self-consistent field. We have controlled that this method
gives same results as an iterative diagonalization of $\hat{h}+\lambda \hat{r%
}^2.$ From a practical point of view, we have done the simulation with up to 
$40$ orbitals. We have chosen a step of $\Delta r=0.2fm$ in r-space.

After having initialized the system, we let it evolve according to the TDHF
equation written in r-space(\ref{eq:evol}). During the expansion of the
system, occupation numbers are kept fixed, implying the conservation of the
entropy of the total system. Note that, in the following, we will sometime
study only part of the system; for such a sub-system the evolution will not
be isoentropic anymore.

We have performed simulations on a lattice of size $480fm$, this size is
big enough to avoid bouncing of particle against the wall. For the
evolution, we have taken a step in time of $0.75fm/c$. Before presenting
results, we will present in an analogous manner the method we have used for
the semi-classical simulation.

\section{Semi-classical model}

The equivalent semi-classical equations are obtained by taking the first term
of the $\hbar $ expansion of Wigner transform of the TDHF equation. This
leads to the Vlasov equation 
\begin{equation}
\frac{\partial f(\vec{r},\vec{p},t)}{\partial t}=\left\{ h[f],f\right\}
\label{eq:Vlasov}
\end{equation}
where $\{.,.\}$ are the usual Poisson brackets and $f(\vec{r},\vec{p},t)$
and $h[f]$ are respectively the one-body phase-space density and the
associated self-consistent mean-field. One of the methods widely used
for solving the Vlasov equation is the lattice Hamiltonian method developed
in ref.\cite{Pan1}.

In this method, the r-space is discretized into a three-dimensional cubic
lattice. Each nucleon is represented by a number $N_{test}$ of ''test
particles''. The density is calculated on each site of the lattice using the
following algorithm: considering a site ''i'', each particle contained in
the neighboring cubes participates to the density at the point ''i'' with a
given weight which depends on the distance between the particle and the
considered site. Formally, this is obtained through the convolution 
\begin{equation}
\bar{\rho}=\rho \otimes G  \label{eq:convol_1}
\end{equation}
where the unfold density is simply given by the density of point-like
test-particles 
\begin{equation}
\rho (x_i,y_i,z_i)=\frac 1{N_{test}}\sum_{n=1}^{AN_{test}}{\ \delta
(x_i-x_n)\delta (y_i-y_n)\delta (z_i-z_n)}  \label{eq:density}
\end{equation}
and where the weight function is given by a distribution of the
density over 2$m$ neighboring sites.

\begin{eqnarray}
G(x,y,z) &=&g(x)\;g(y)\;g(z)  \label{eq:G} \\
g(\alpha ) &=&\frac{(m\Delta r-\left| {\alpha }\right| )}{(m\Delta r)^2}%
\Theta (m\Delta r-\left| {\ \alpha }\right| )  \label{eq:g}
\end{eqnarray}
In this expression, $\Theta $ corresponds to the Heavyside function, so that 
$(m\Delta r)$ corresponds to the range of the smoothing used to compute the
density. Finally we get the following density at each site $i$ of the
considered lattice 
\begin{equation}
\bar{\rho}(x_i,y_i,z_i)=\frac 1{N_{test}}\sum_{n=1}^{AN_{test}}{\
g(x_i-x_n)g(y_i-y_n)g(z_i-z_n)}  \label{eq:convol_2}
\end{equation}
This density is used to compute the energy functional $E\left[ \bar{\rho}%
\right] $ (see eq. (\ref{eq:E})) at each point $i$ of the lattice, in order
to define the Hamiltonian ${\cal H}$ of the system \cite{Pan1} 
\begin{equation}
{\cal H}(\vec{r}_n,\vec{p}_n)=\sum_n\frac{p_n^2}{2m}+\sum_i\Delta r^3E\left[ 
\bar{\rho}\right] \left( \vec{r}_i\right) +E_{Coulomb}  \label{eq:H}
\end{equation}
where, following our conventions, the sum over $n$ runs over all the
test-particles while the sum over i corresponds to the sum over lattice
sites. The evolution of the system consists in the classical Hamilton
equations for each test particle $n$ which have been solved using the
standard leap-frog algorithm 
\begin{eqnarray}
\vec{r}(t+\Delta t) &=&\vec{r}(t-\Delta t)+2\Delta t\;\frac{\vec{p}(t)}m,
\label{eq:evol_r} \\
\vec{p}(t+2\Delta t) &=&\vec{p}(t)+2\Delta t\;\vec{F}(t+\Delta t),
\label{eq:evol_p}
\end{eqnarray}
where the force is calculated according to 
\begin{equation}
\vec{F}=-~\frac{\partial {\cal H}(\vec{r}_n,\vec{p}_n)}{\partial \vec{r}_n}
\label{eq:F}
\end{equation}
Using the expression (\ref{eq:H}) for the Hamiltonian ${\cal H}$, it is
easy to demonstrate that the force $\vec{F}$ derives from a potential $\bar{U%
}$ which is related to the mean field potential $U\left[ \bar{\rho}\right] $
through the relation 
\begin{equation}
\bar{U}(x,y,z)=\frac 1{(2m)^3}\sum_iU(x_i,y_i,z_i){\ }  \label{eq:convol_U}
\end{equation}
where the sum runs over the $2m$ neighboring cells of the position $(x,y,z).$

This method has demonstrated its accuracy and is now widely used in
theoretical description of Heavy-Ion collision \cite{Pan1}. We have thus
used it and compared results with quantum simulation. It should be noticed
that other methods are reported in the literature using either slightly
different lattice discretization of the density or gaussian smoothing in
order to fold the point-like test-particle density into a smooth
nucleon density.
Therefore, all the semi-classical methods contain a smoothing procedure in
order to generate the nucleon density used in the energy functional. The
presented results do not depend on the particular numerical implementation
of this smoothing.

\smallskip In practice, we have adopted the value $N_{test}=200$ which
corresponds to a rather large number in such type of simulations. For the
Vlasov evolution, we have taken the three dimensional code TWINGO developed
in ref.\cite{Sur92,Gua1}. The simulation are done with a range for the convoluting
function $g$ of $2fm$, the lattice grid has a step of $1fm$ and the step in
time is the same as in the quantum simulation: $0.75fm/c$. The total r-space
is a box of size $40\times 40\times 40$, when a particle reachs the boundary,
it is not evolved anymore.

Before describing the results, we will get more insight into the simulation
and in particular into the initialization procedure.

\subsection{Initialization and stable states of the Vlasov dynamics.}

\subsubsection{\protect\smallskip Constrained Thomas-Fermi solution}

The first point is to define stable spherical nucleus at temperature $T$
under the constraint $\lambda r^2$. For this purpose, we have only to
compute the spherical one-body phase space density $f(r,p)$ of the
Thomas-Fermi type 
\begin{equation}
f(\vec{r},\vec{p},T)=\left( \exp \left( \frac{\vec{p}^2/2m+\bar{U}(\vec{r}%
)+\lambda r^2+V_{Coulomb}-\mu }T\right) +1\right) ^{-1}  \label{eq:f}
\end{equation}

\noindent which indeed maximizes the entropy 
\begin{equation}
S=4\idotsint \frac{d^3r\;d^3p}{h^3}\left( f\log f+(1-f)\log (1-f)\right)
\label{eq:S_class}
\end{equation}
under the various constraints. In 
expression (\ref{eq:f}), $\mu $ is the chemical potential 
and $\bar{U}(\vec{r})$ is the
potential consistent with the lattice Hamiltonian method (see Eq. (\ref
{eq:convol_U}))which is obtained using the folded density $\bar{\rho}$. Note
that, in order to speed up the initialization process, we have used the
spherical symmetry: we have replaced Eqs. (\ref{eq:convol_2}) and (\ref
{eq:convol_U}) by convolutions with a spherical function 
\begin{eqnarray}
\bar{\rho}(r) &=&\int_{r=0}^{r^{\prime }=+\infty }\rho (r^{\prime
})\;g(r^{\prime }-r)4\pi r^{\prime }{}^2dr^{\prime }  \label{eq:conv_sp_rho}
\\
\bar{U}(r) &=&\int_{r=0}^{r^{\prime }=+\infty }U\left[ \bar{\rho}\right]
(r^{\prime })\;g(r^{\prime }-r)4\pi r^{\prime }{}^2dr^{\prime }
\label{eq:conv_sp_U}
\end{eqnarray}
where the function $g$ is a function analogous to the expression (\ref{eq:g}%
) 
\begin{equation}
g(r)=\frac{(r_0-r)}{(r_0)^2}\Theta (r_0-r)  \label{eq:g_sp}
\end{equation}
but with a range $r_0$ adapted to provide the same averaged squared-radius
as $G$.

The numerical procedure used to build a constrained Thomas-Fermi distribution in
phase space can be separated in two steps:

\begin{itemize}

\item  the first one consists in an iterative definition of $f(r,p,T)$:
starting with a potential (at the first iteration the potential is assumed
to be a Wood-Saxon), we calculate the phase space density according to
expression (\ref{eq:f}) using a chemical potential $\mu $ which conserves
the particle number. Then, we obtain the density $\rho (r)$ by integrating $%
f(r,p,T)$ over $p$ and we calculate the folded density $\bar{\rho}(r)$ and
the associated potential $\bar{U}(r)$ using Fast Fourier transform
techniques. Finally, we iterate the operation until a full convergence is
reached. This provides us with a continuous self-consistent phase-space
density $f(r,p,T).$

\item  The second step consists in distributing test particles according to
the phase space distribution, $f(r,p,T)$ using a 
standard Metropolis algorithm \cite{Met1}. 
\end{itemize}

The stability over time of a nucleus initialized in its ground state
demonstrates the accuracy of this method in finding stable solutions in
lattice Hamiltonian framework.\footnote{%
Indeed, we have not observed the emission of a single test-particle during
the typical simulation time used in the following.(more than $300fm/c$)}

\subsubsection{Scaling procedure}

As already discussed in the quantum case, we have also studied a second
class of initial conditions namely those obtained through a scaling.
In order to follow the approximation made in the quantum simulation, the
mean-field potential, included in the definition of the finite temperature
phase space density (eq. (\ref{eq:f})), has been approximated by the
zero-temperature one (including a small constraining external potential $%
\lambda r^2$, $\lambda =0.25MeVfm^{-2}$). For each of these initial
conditions, the temperature is chosen in order to get the expected entropy (%
\ref{eq:S_class}).Then, the test-particle distribution is generated using
the Metropolis algorithm as discussed above. Finally the momenta and the
positions of the test particles are scaled according to eq. (\ref
{eq:scaling1}). Since this scaling preserves $f$ and the phase space volume $%
d^3rd^3p/h^3, $ it does not change the entropy (\ref{eq:S_class}). We are
thus generating a family of isoentropic initial conditions.

\smallskip

\smallskip This scaling procedure can be also used to construct an ensemble
of initial conditions which can be considered as isothermal by changing the
temperature $T$ in order to compensate the effects of the dilution (or the
compression) on the temperature prior to the scaling according to Eq. (\ref
{eq:T_scale}).

\subsection{Discussion on smoothing in semi-classical methods}

In this section, we will discuss the possible bias of the comparison between
quantum and semi-classical simulations due to the test-particle method.
Indeed, in any semi-classical approach based on a test-particle method, a
smoothing of the phase space density should be included in order to compute
the mean-field potential. As already mentioned in \cite{Eric}, this
smoothing acts as an effective range for the forces. Indeed, considering the
density of particles $\rho $, the actual numerical smoothed density $\hat{%
\rho}$ is given by the folding product with a smooth function $g$

\smallskip 
\begin{equation}
\hat{\rho}=\rho \otimes g
\end{equation}

\noindent The function $g$ used in our simulations is given by equation (\ref
{eq:density}) but $g$ could take other expression. Expending the Fourier
transform of $g$ at the second order in $k$ this folding product can be
approximated by

\begin{equation}
\hat{\rho}=\rho -\beta \Delta \rho
\end{equation}

\noindent where $\beta $ is the range of the function g. Then, the
time-dependent mean-field becomes 
\begin{equation}
U\left[ \hat{\rho}\right] =\frac 34t_0\rho +\frac{(2+\sigma )}{16}t_3\rho
^{1+\sigma }-\beta \Delta \rho \left( \frac 34t_0+\frac{(2+\sigma )(1+\sigma
)}{16}t_3\rho ^\sigma \right)  \label{Eq:10}
\end{equation}
\noindent where one can recognize a typical finite range term in the last
part of Eq. (\ref{Eq:10}). However, it should be noticed that the above
expression corresponds to a very specific surface contribution.

As a direct consequence, due to this additional surface energy, we will see
that energies calculated in semi-classical approximations will be greater
that those calculated in quantum simulations. This might seems surprising
since in absence of quantum correction terms in the semi-classical
potential, one would expect the surface energy present in the semi-classical
picture to be smaller than the quantum equivalent (or even zero). In fact,
this surface energy is only generated by the numerical method. Indeed, in
the quantum case, since wave functions are not localized, no smoothing is
needed and the algorithm does not introduce spurious additional surface
term. However, it is well known that even with a zero range force with no
momentum dependence (to simulate a finite-range), a quantum calculation will
exhibit non trivial surface properties due to the role of the kinetic term
and because of the self consistence effects on the wave functions.
Conversely, a semiclassical calculation with a zero range force should
present sharp edges (at zero temperature). 
Then, surface properties, in semi-classical
calculation, are influenced by the smoothing
procedure.

\smallskip This difference in the surface properties can be seen as a major
drawback of semi-classical approaches. 
Of course one might think to take advantage of the
presence of this numerical finite range in semi-classical pictures in order
to mimics some quantum effects or some surface terms in the considered
interaction. In the next chapter, we will more carefully investigate this
possibility, however, we can already mention some difficulties:

\begin{itemize}
\item  It is never easy to rely on the numerics in order to mimic physical
effects.

\item  One might try to select the numerical coarse-graining, but normally,
it is fixed by some other arguments such as numerical accuracy (see Eq. (\ref
{Eq:10})).

\item  since the surface effects are coming from the folding procedure in
the mean-field potential their importance is fixed by the mean field
strength (see Eq. (\ref{Eq:10})). Since this strength is density dependent,
the surface effect will have the same dependence. This might be not
realistic since they are supposed to mimic surface effects of various origin
and so with various density dependence: interaction range, quantum kinetic
term, finite range of wave functions and self-consistent effects. This point
is partly discussed in ref.\cite{colo}. 

\item  In any realistic cases, where surface term exists in the interaction,
one should be careful in order to not double count these finite size
effects.
\end{itemize}

\smallskip


We observe that, a semi-classical simulation, which is supposed to be based
on a given force (\ref{eq:inter}), includes a spurious additional
interaction (\ref{Eq:10}). As we will see, this will contribute
to the observed differences with the quantum simulation associated to 
(\ref{eq:inter}).
However, realistic interactions usually include a finite range term, so
that it might be interesting to investigate the possibility to mimics
quantum dynamics using a finite range interaction with a semi-classical
approach eventually taking into account the numerical surface term in the
potential.

\subsection{ Discussion of surface effects.}

In this section we consider a mean-field potential taking
into account a finite range effect in the TDHF mean-field\footnote{%
In the skyrme force, the parameter $c$ is related to the parameters $t_1$
and $t_2$\cite{Vau2,uka}.}.

\begin{equation}
U(\rho )=\frac{3}{4}~t_{0}~\rho +\frac{(\sigma +2)}{16}~t_{3}~\rho ^{\sigma
+1}+c~\nabla ^{2}\rho .  \label{Skyrme}
\end{equation}

Usually, in the TDHF
calculation, the Laplacian terms is simulated by a finite range potential of
the Yukawa type in order to avoid well-known numerical instabilities\cite
{uka}. Therefore, the mean-field potential has been computed using the
relation

\begin{equation}
U(\rho )=\frac{3}{4}~t_{0}^{^{\prime }}~\rho +\frac{(\sigma +2)}{16}%
~t_{3}~\rho ^{\sigma +1}+V_0~\rho \otimes Y.  \label{Eq:11}
\end{equation}

\noindent where $Y$ is a Yukawa folding function 
\begin{eqnarray}
Y(r)&=&\frac{\exp(-|r|/a)}{|r|/a}.  \label{yukawa}
\end{eqnarray}

\noindent and where $t_{0}^{^{\prime}}$ is function of $a,V_0$ and the $t_0$
used in the classical calculation. Indeed, expanding the Fourier transform
of the folding product up to the second order in $k$ we can derive the
following expression for the potential

\begin{equation}
U(\rho )=\frac{3}{4}~\underbrace{ (t_{0}^{^{\prime }}+\frac{16\pi V_0 a^3}
{3})}_{t_0}~\rho +\frac{(\sigma +2)}{16}~t_{3}~\rho ^{\sigma +1} +%
\underbrace{2V_0\pi a^5}_{c}~\nabla ^{2}\rho.  \label{pot3}
\end{equation}

For numerical applications, we have chosen the same values of $t_0$ and $t_3$
as before in the semi-classical simulation. For the quantum case, we have
chosen for $a$ and $V_0$ respectively $0.37$ fm and $-200$ MeV 
(which gives a $%
t_0^{^{\prime }}$ value of $830.3$ MeV fm$^3$) 
in order to have almost the same energy
for the ground state in quantum and in semi-classical case at zero
temperature. In the following, this particular simulation will be referred
as ''{\it Quantum+surface}''.

\section{Results}

In this section we will focus on a detailed comparison of the two considered
approaches, the semi-classical Vlasov dynamics and the full quantum TDHF
treatment (with and without additional surface terms). We will take a hot
and compressed or diluted Double magic $^{40}$Ca nucleus as a test case.


\subsection{Initialization of nuclei under extreme conditions}


In this section, we will illustrate different initial conditions of
compression and temperature. We will first discuss density profiles obtained
after the initialization step and then, present the various excited nuclei we
are considering in the evolution.

\subsubsection{Density profiles}

\begin{figure}[tbph]
\begin{center}
\includegraphics*[height=10cm,width=15cm]{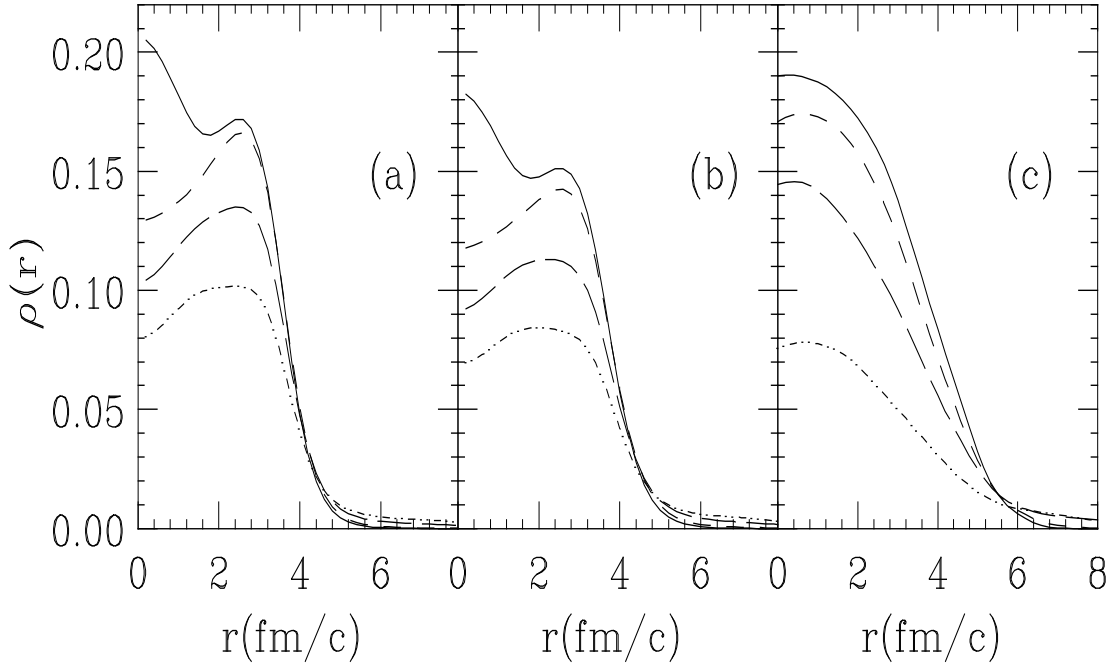}
\end{center}
\caption{Initial density profiles of a $^{40}Ca$ nucleus at various
temperatures for a constraining field $\lambda r^{2}$ with $\lambda =0.25$
MeV/fm$^{2}$: a)Quantum approach; b) Quantum+surface c)Semi-classical. In
each figures are displayed temperature $T=0$MeV(solid line), $5$MeV (dashed
line), $10$MeV (long-dashed line), $15$MeV (short-dashed line).}
\label{fig:1}
\end{figure}

In figure (\ref{fig:1}), we compare quantum and semi-classical constrained
mean-field solutions obtained at various temperature ($T=0$, $5$, $10$ and $15$
MeV) with a small constrained field $\lambda r^2$ with $\lambda =0.25 $ MeV/fm%
$^2$. As expected, the semi-classical simulation, which, in addition to the
smoothing procedure, does not include shell effects, presents smoother
initial density profiles. Directly connected to the smoothing, the quantal
density profiles always present a smoother surface than the semi-classical
one. In addition, the Root Mean Square radii (RMS) and the central
densities are reported in table \ref{table:RMS}. We observe that these
quantities are slightly different in semi-classical case compared to both
quantum simulations. In particular, the temperature dependence of such
global observables indicates a different response of the nuclei to the
heating process within the two approaches.


\smallskip 
\TeXButton{B_tab}{\begin{table}[tbp] \centering} 
\begin{tabular}{|c|c|c|c|c|c|c|}
\hline
& \multicolumn{2}{|c}{\it Quantum} & \multicolumn{2}{|c}{\it Quantum} & 
\multicolumn{2}{|c|}{\it Classical} \\ 
& \multicolumn{2}{|c}{} & \multicolumn{2}{|c}{\it +Surface} & 
\multicolumn{2}{|c|}{} \\ \cline{2-7}\cline{2-7}
& RMS & $\rho $ & RMS & $\rho $ & RMS & $\rho $ \\ \hline\hline
$T=0MeV$ & 3.14 & 0.172 & 3.31 & 0.153 & 3.09 & 0.196 \\ \hline
$T=5MeV$ & 3.30 & 0.148 & 3.53 & 0.132 & 3.36 & 0.179 \\ \hline
$T=10MeV$ & 4.30 & 0.130 & 4.65 & 0.109 & 4.93 & 0.144 \\ \hline
$T=15MeV$ & 5.55 & 0.098 & 5.94 & 0.082 & 8.03 & 0.080 \\ \hline
\end{tabular}
\caption{Root Mean Square radius (RMS) and central
densities of a $^{40}$Ca nucleus initialized with a constraining field
 $\lambda r^{2}$ with $\lambda =0.25$ MeV/fm$^{2}$
at various temperatures
\label{table:RMS}}\TeXButton{E_tab}{\end{table}}

\smallskip 
In the scaling procedure, for isoentropic initial conditions, density
profiles of compressed or dilated nuclei will be obtained by a
scaling of those displayed in Fig.\ref{fig:1}. 


\subsubsection{Isoentropic initial conditions}

Important information about a system is provided by 
the total energy $E$ as a function of $\rho $ for various entropies. When
considering infinite nuclear matter this is the usual Equation Of state
(EOS), $E=E(\rho ,S)$ as shown in fig. 2 for the considered interaction. 
This EOS was extensively used in
order to understand the properties and the dynamics of nuclear matter. 
%
For each entropy, the minimum of the energy corresponds to a stationary
isoentropic
solution. 
Furthermore, the second derivative of the energy with respect to $\rho $
around the stationary point is directly related to the frequency 
of the monopole oscillation around this point. 
The isoentropic spinodal region is defined as points where this curvature is
negative or equal to zero. 
This region is of particular interest since one of the
possible scenarii of multifragmentation is that a highly excited nucleus
could enter this region and become unstable against the partition of the
matter in fragments. 
\begin{figure}[tbph]
\begin{center}
\includegraphics*[height=13cm,width=15cm]{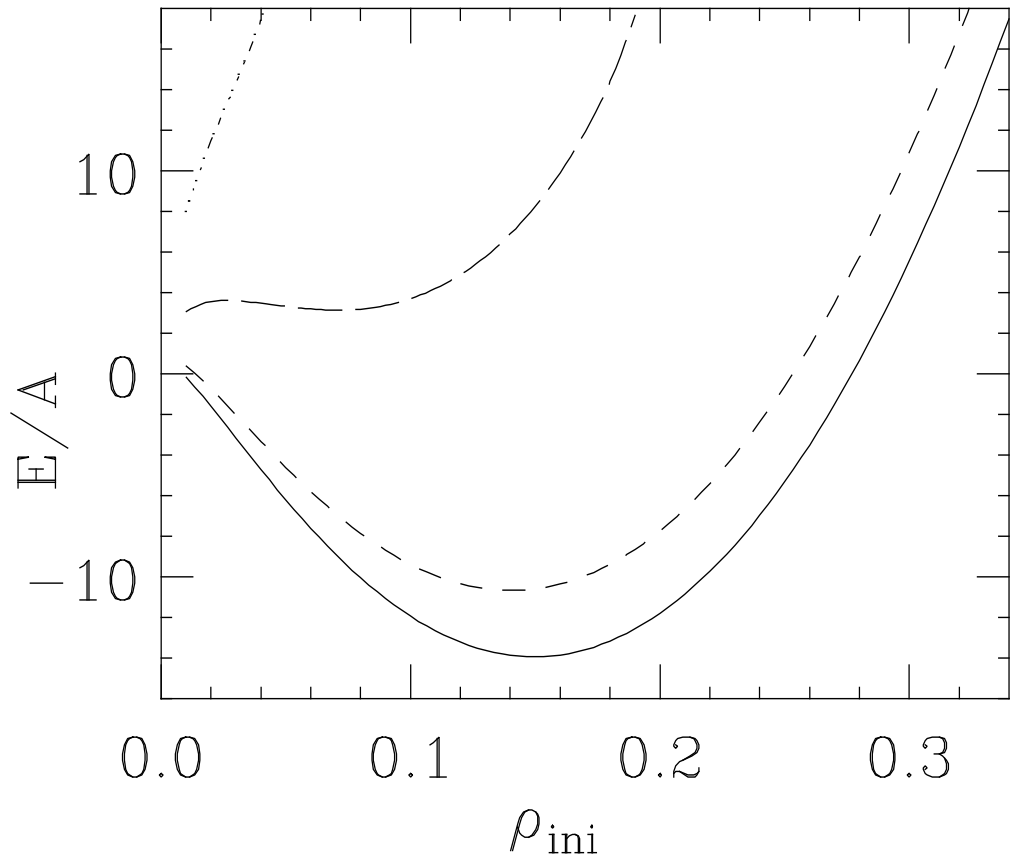}
\end{center}
\caption{Binding energy of the infinite nuclear matter as a function of its
density, the curves correspond respectively to $S/k_B=0,~1.1,~2.35$ and $3.28
$.}
\label{fig:infinite}
\end{figure}

\begin{figure}[tbph]
\begin{center}
\includegraphics*[height=9cm,width=16cm]{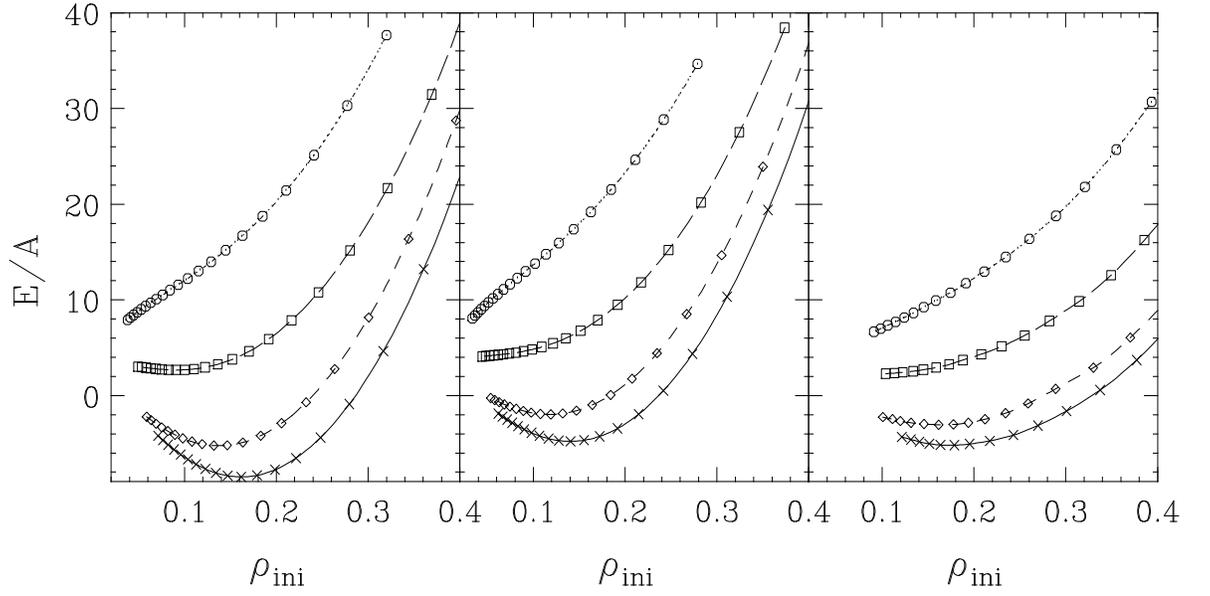}
\end{center}
\caption{Binding energy of a hot and compressed/diluted Ca nucleus as a
function of its central density, the curves corresponds respectively from
bottom to top to $S/k_B=0~(\times),~1.1~(\diamond),~2.35~(\Box)$ and$%
~3.28~(\circ)$ on each figures. Left: Quantum (without surface). Middle:
Quantum+Surface. Right: Semi-classical.}
\label{isent:1} 
\end{figure}

Let us built an equivalent of the EOS for a finite system. In order to do
so, we can plot the relation between the central density $\rho $ and the
energy per particles $E/A$ of the system for the various initial entropies
considered. In Fig.\ref{isent:1}, we report four different curves which
correspond to entropies $S/k_B=0,~1.10,~2.35$ and $3.28.$ These curves are
obtained by applying the general scaling procedure to each models:
occupations numbers are kept fixed while different compressions/dilations are
considered. 

Looking first at the minimum of the curve $S/k_B=0,$ we see that the
semi-classical stationary point has an energy greater than that of the
quantum system (without surface).This is a direct consequence of the
smoothing procedure. The additional surface term gives an extra energy of
the order of $3$ MeV. Note that, in the second quantal calculation (middle
part), the surface term was adjusted in order to have the same ground-state
energy as in the semi-classical case. If we now focus on the curvature of
isoentropic curves, we see that they appear smaller in
semi-classical than in both quantal calculations. Note that, if
we suppose an evolution at contant entropy and energy, 
these curvatures are direcly related to the response to a monopolar excitation.
In such a context, according to figure \ref{isent:1}, 
the semi-classical model should have a larger amplitude than quantal
one. However, in dynamical simulations, physical systems
have neither constant energy nor constant
entropy in particular due to particle evaporation. As we will show later,
the difference in evaporation processes in addition to the damping 
due to the use of 
test-particle method will strongly reduce the amplitude of monopolar
vibrations in
semi-classical simulations making it smaller than the quantum analogue.

\begin{figure}[tbph]
\begin{center}
\includegraphics*[height=10cm,width=10cm]{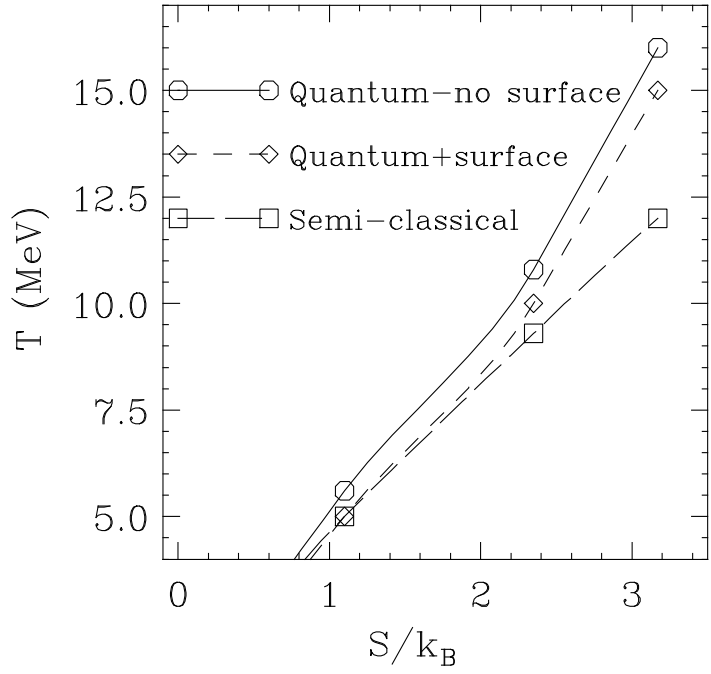}
\end{center}
\caption{Entropy as a function of the temperature for a $^{40}$Ca nuclei
obtained when a small external field $\lambda r^2$ with $\lambda =0.25$
MeV/fm$^2$ is applied. Solid line: Quantum (without surface), dashed line:
Quantum+surface, long-dashed line: Semi-classical.}
\label{fig:ST}
\end{figure}

Finally, we want to point out that temperature creates more dilution and
disorder in semi-classical calculation than in quantum one as we already
noticed in Fig.\ref{fig:1}. We can quantify this more easily considering for
instance the dependence of the entropy with temperature at a fixed
compression factor. In Fig. \ref{fig:ST}, we report such a dependence for
the three models when no compression is applied. We see in particular, that
higher temperatures are necessary in quantal cases in order to create the
same entropy than in a semi-classical simulation. This is a great difference
between quantum and classical treatment of nuclei.

\subsection{\protect\smallskip Monopole oscillations and thermal expansion}

\subsubsection{Examples of expansion}

\begin{figure}[tbph]
\begin{center}
\includegraphics*[height=16cm,width=16cm]{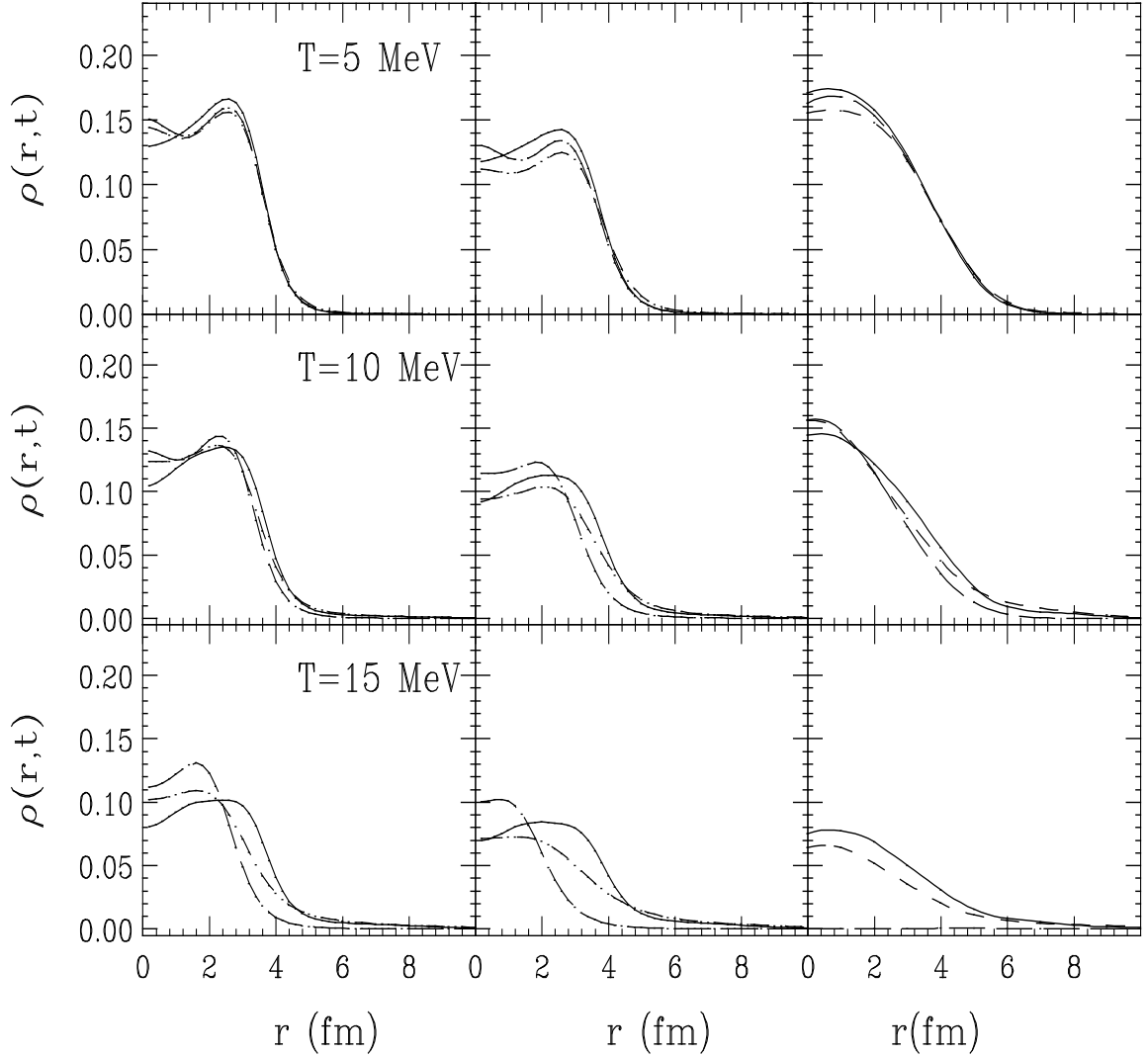}
\end{center}
\caption{Density profiles for various initial temperatures - for $T= 5$, $10$%
, $15$ MeV (from the top to the bottom) - and for: Left part: Quantum
(Without surface). Middle Part: Quantum+Surface. Right part: Semi-classical.
In each figure various curves corresponds to different times: Solid line: $%
t=0$ fm/c. Dashed line: $t=30$ fm/c. Long-dashed line: $t=300$ fm/c.}
\label{fig:2}
\end{figure}

Let us now consider the time evolution of the density profiles presented in
figure \ref{fig:1} for the constrained method. 
In figure \ref{fig:2}, we present the evolution of a hot system for
different temperature $T=5,~10,~15~MeV$. 
From a qualitative point of view, the semi-classical and the TDHF evolutions
are in reasonable agreement except at high temperature, for which,
the semi-classical nucleus does not resist to the heat and
to vaporizes easily than its quantum analogue. In fact, not only the
vaporization process appears different and seems to start at a lower value
of the initial temperature but also the whole particle evaporation is faster.

\smallskip

Before investigating in more details this observation and before discussing
the possible reasons of this behavior let us first study
quantitatively the various observed differences between the semi-classical
and the quantum approaches.

\subsubsection{Illustration of the monopole oscillation}

Let us first study the collective monopole motion associated with a
compressed or diluted system.

\begin{figure}[tbph]
\begin{center}
\includegraphics*[height=16cm,width=12cm]{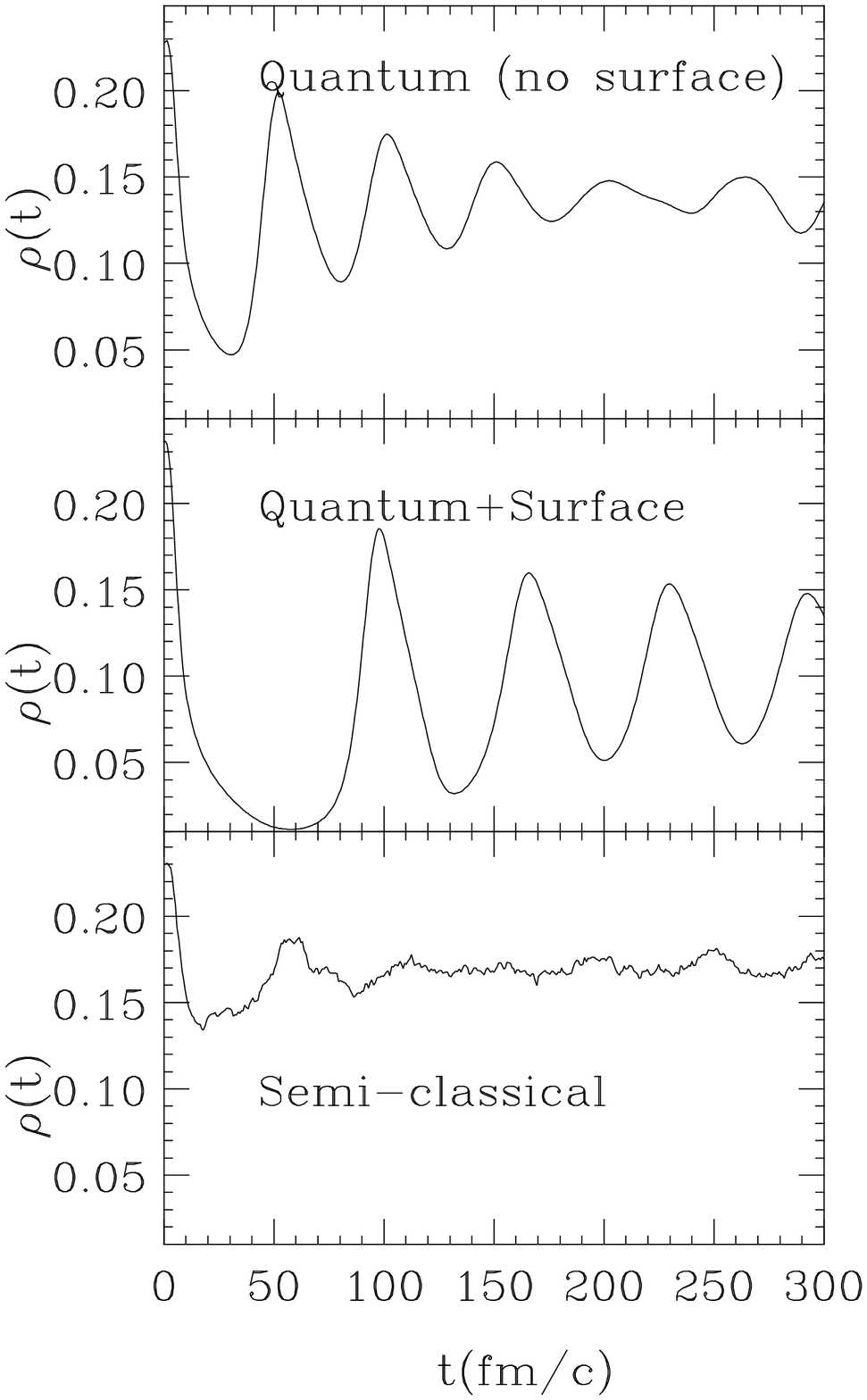}
\end{center}
\caption{Illustration of oscillation of the central density in time for
different models. In all models, initial conditions correspond to a $^{40}$%
Ca nuclei at $T=5$ MeV and an initial density $\rho_{ini} =0.23$ fm$^{-3}$.
Top: Quantum (Without surface). Middle: Quantum+surface. Bottom:
Semi-classical. }
\label{fig:3}
\end{figure}

\smallskip

Figure \ref{fig:3} shows the central density as a function of time, predicted
by the quantum evolution for a heated and compressed Ca nucleus initialized
at $T=5$ MeV and with an initial density $\rho_{ini} =0.23$ fm$^{-3}$. In
all cases, the central density presents an oscillating pattern due to the
monopole vibration of the nucleus. In addition, in Quantum simulation
beatings of different modes which are characteristic of the Landau spreading
of the breathing mode onto various components\cite{mono,Lif}%
, are observed even at high temperature (see Fig. \ref{fig:3} (Top)). 
Note also, that the central density in semi-classical
simulations , exhibits an important noise added on top to the monopole
vibration. These fluctuations are due to the finite number of
test-particles. In order to quantitatively study this vibration, for
different initial conditions, we have considered two observables as a
function of the initial entropy and density:

\begin{itemize}
\item  the value of the minimum of the central density reached during the evolution
(noted $\rho _{\min }$
which provides a measure of the amplitude of the monopole vibration;

\item  the time needed to reach this minimum (noted $t_{\min }$)which, in
case of a pure oscillatory motion, is directly related to half of the period
of the vibration.
\end{itemize}

\smallskip

In the following, we will present results where events are grouped according
to their entropies. 

\subsubsection{Expansion of isoentropic initial conditions}


\begin{figure}[tbph]
\begin{center}
\includegraphics*[height=9cm,width=16cm]{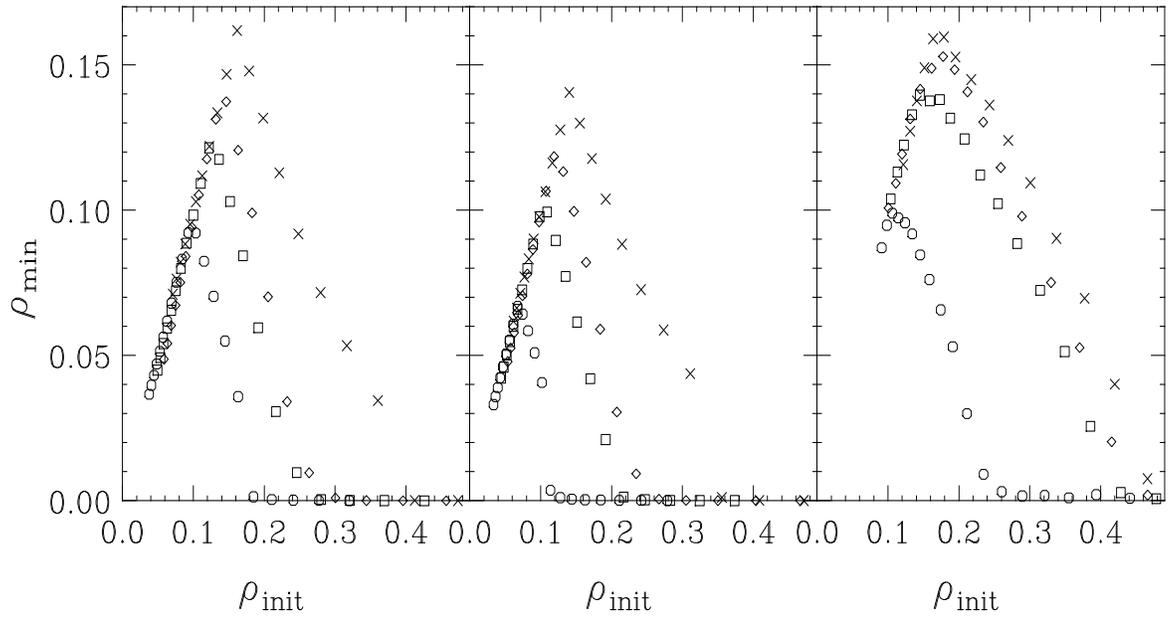}
\end{center}
\caption{The central density at the maximum of the dilution (turning point)
as a function of the initial density for a $^{40}$Ca nucleus at various
initial temperatures (from top to bottom points corresponds respectively to $%
S/k_B=0~(\times),~1.1~(\diamond),~2.35~(\Box)$ and$~3.28~(\circ)$). Left:
Quantum (without surface). Middle: Quantum+Surface. Right: Semi-classical
simulation.}
\label{fig:roi_romin2}
\end{figure}

The value of the first minimum of the central density oscillations ($\rho
_{\min }$) is displayed in figure (\ref{fig:roi_romin2}) as a function of
the value of the initial density ($\rho _{init}$). For a given entropy, this
curve presents two branches. The first one (low initial density region) 
corresponds to the case of
a minimum density equal to the initial one, i.e. the nucleus is already
diluted compared to the stationary initial condition. Therefore this first
branch corresponds to $\rho _{min}=\rho _{init}$. The second branch gives
the amplitude of the collective motion, i.e. the maximum dilution, as a
function of the initial compression. The point where these two branches met
is the stationary point for which no oscillations are observed. Note also
that at finite temperature even if the system is stationary, it cools down
by particle evaporation.

\smallskip

\begin{figure}[tbph]
\begin{center}
\includegraphics*[height=10cm,width=10cm]{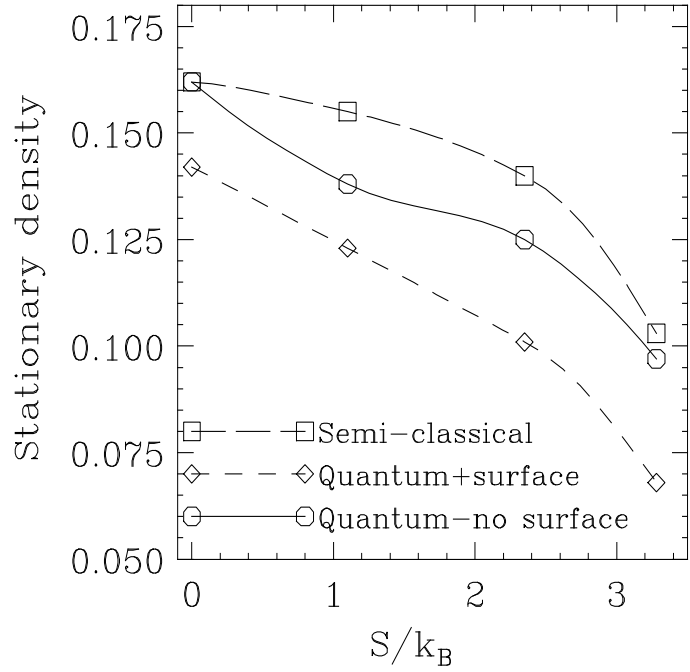}
\end{center}
\caption{Stationary points obtained from the ($\rho _{min},\rho _{init}$)
diagram. The stationary density value is plotted as a function of the
entropy. Solid-line: Quantum (without surface). Dashed-line:
Quantum+Surface. Long dashed line: Semi-classical.}
\label{fig:station}
\end{figure}

Comparing now the classical and quantal simulations, one can see several
differences. First of all, it is clear that the effect of entropy (or
temperature) is weaker in semi-classical simulations when the excitation is
not too high. In particular the stationary points do not exhibit the same
entropy (or temperature)
dependence in the quantum case or the classical case
(see. Fig. \ref{fig:station}).

\begin{figure}[tbph]
\begin{center}
\includegraphics*[height=10cm,width=10cm]{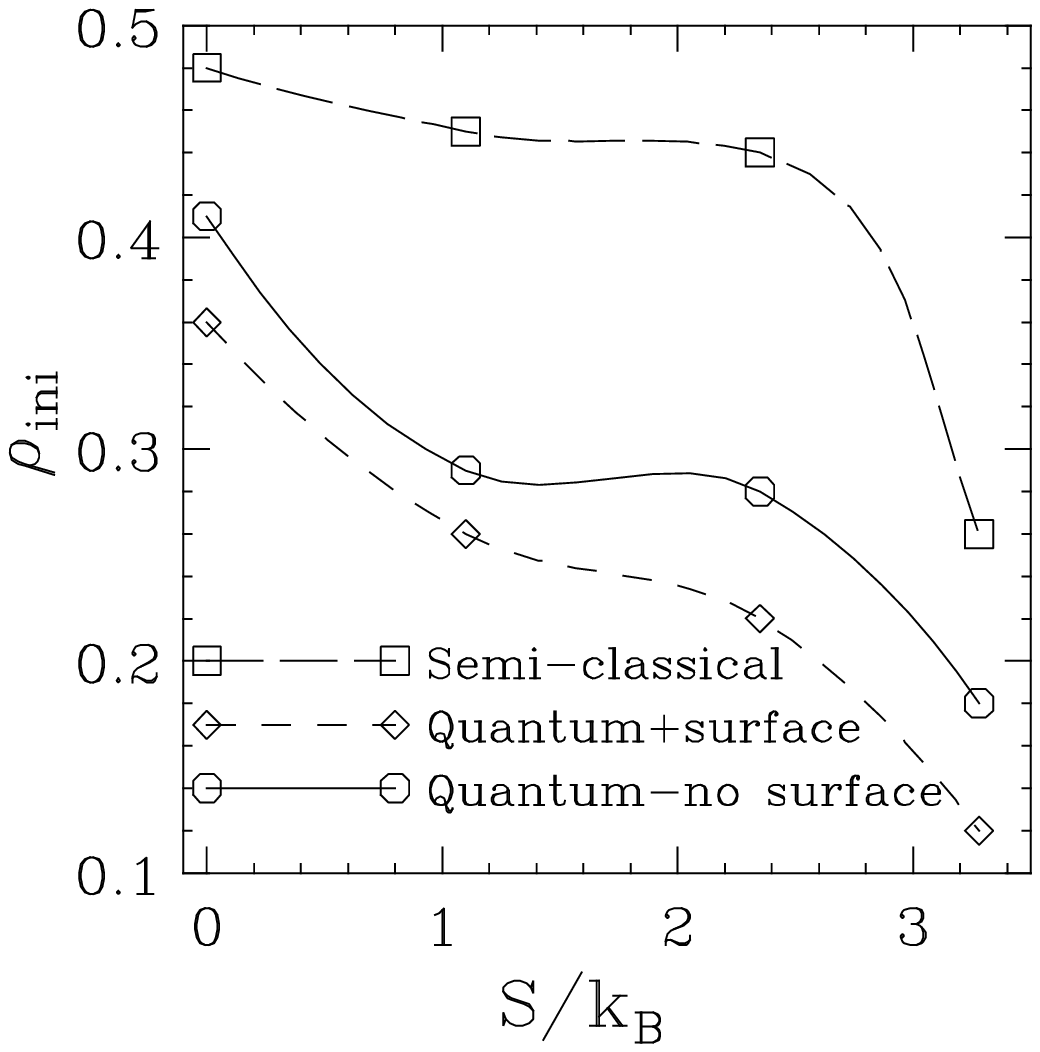}
\end{center}
\caption{Initial density for which the central density reach $0$ during the
evolution (from the ($\rho _{min},\rho _{init}$) diagram). The initial
density value is plotted as a function of the entropy. Solid-line: Quantum
(without surface). Dashed-line: Quantum+Surface. Long dashed line:
Semi-classical.}
\label{fig:vap}
\end{figure}

In a similar manner, the initial densities for which the central density
will eventually reach 0 (i.e. either the nucleus 
vaporizes or leads to an hollow system)
have different temperature dependence(see Fig.\ref{fig:vap}). 
At low temperatures,
semi-classical approaches are associated with a fast emission of particles.
This faster evaporation as compared with the quantum dynamics induces a
faster damping of the collective motion. Moreover, it is
known that the fluctuations induced by test particle methods as well as the
possible turbulent dynamics of classical fluids generate an additional damping
of collective motions in semiclassical approaches \cite{Eric}%
. All these different dampings reduce the amplitude and the frequency of
vibrations. In such a way, a more compressed nucleus can survive more easily
in 
semi-classical simulations because it can dissipate energy more efficiently than
its quantum analogue and has a longer time to cool down. At high
temperature, all particles are immediately evaporated in the semi-classical
approaches leading to the vaporization of the nucleus.

If we now look at the slopes of $\rho _{\min }$ versus $\rho _{ini}$ in the
right part of figure \ref{fig:roi_romin2}, we can see that the slopes appear
smaller in semi-classical case than in both quantum cases. This behavior
indicates a larger damping of the collective motion together with a softer
response to an external compression of the nuclei in the semi-classical
treatment, compared with the quantum ones. This apparent softness could
appear surprising {\it a priori }since we are considering models with same
EOS for the infinite nuclear matter. However, we have seen in the previous
section that the EOS for finite systems are different in the various
simulations (see Fig. \ref{isent:1}). 

In conclusion, a higher compression is needed in semi-classical treatments
as compared with quantum approaches in order to reach the same minimum
density. This corresponds to a smaller amplitude and a slower collective
motion. Part of this difference could be explained by the smoothing
procedure necessary in semi-classical method to generate the mean-field.
However, adding a surface term in quantum mechanics reduce only a little
this difference, indicating that physical aspect such as, for instance,
dampings, evaporation, time-scale... are different. In the following, we
will quantify differences in time-scale and in the evaporation process.

\subsubsection{Monopole vibration period}

\begin{figure}[tbph]
\begin{center}
\includegraphics*[height=9cm,width=16cm]{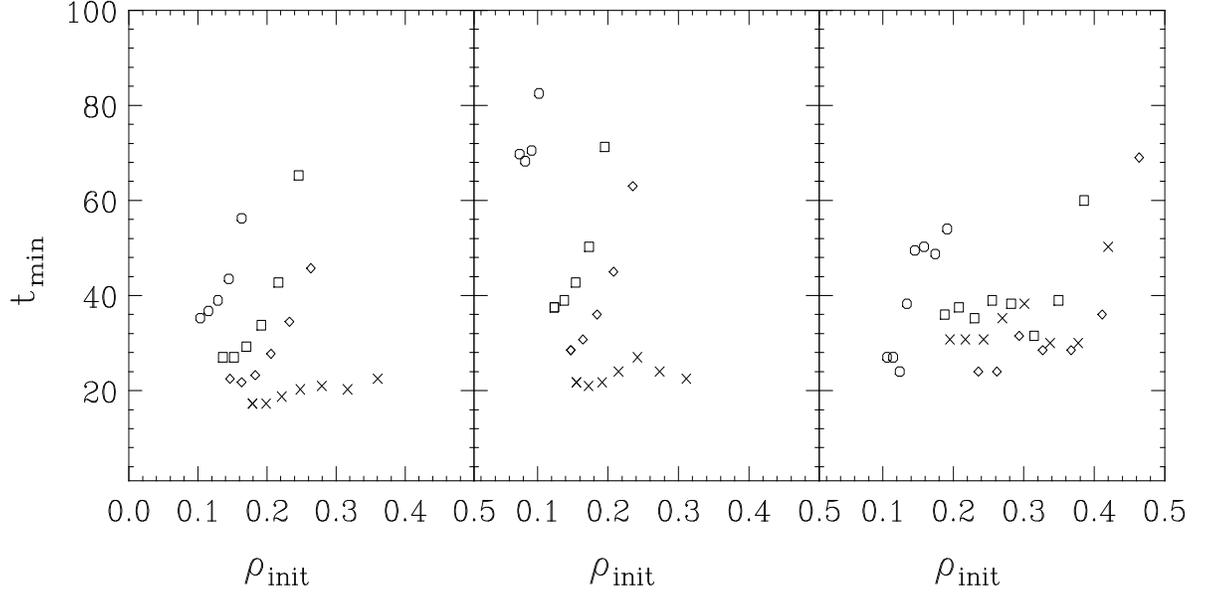}
\end{center}
\caption{Time $t_{\min }$ to reach the minimal density as a function of the
initial density $\rho _{init}$ at various initial entropies $%
S/k_B=0~(\times),~1.1~(\diamond),~2.35~(\Box)$ and$~3.28~(\circ)$. Left:
Quantum (without surface). Middle: Quantum+Surface. Right: Semi-classical
simulation. Note that, near the stationnary point, the time $t_{\min }$
could not be extracted in semi-classical simulation due to the strong
fluctuations induced by test-particle method (see figure 6). These points
are thus not represented in the figure. 
}
\label{fig:time_s}
\end{figure}

In the previous section, we have mentioned that semi-classical methods lead
to much softer response of the nucleus. As a direct consequence, we could
expect a slower expansion dynamics in the semi-classical treatment.
In Fig. \ref{fig:time_s}, we have
plotted the time (noted $t_{\min }$) necessary to reach the lowest central
density $\rho _{\min }$ for isoentropic initial conditions. At high
initial compression, this time goes to infinity since the system is
vaporized. For initial dilatation below the stationary point, this time is
simply equal to zero(or nearly zero) since the system immediately contracts.
Inbetween, we have finite, non-zero $t_{\min }$ which could be interpreted
to be one-half of the period of monopole vibration. In the two quantum
cases, at temperature or entropy equal to zero, the monopole vibration
period is close to $80$ fm/c and appears to be almost independent of the
initial perturbation (which is similar to the harmonic
picture). In the classical case, the period of monopole vibration is of the
order of $120$ fm/c. This indicates again a softer response of the classical
system accompanied with a longer time of expansion. 
If we now look to higher temperature (or entropies), we see that in
quantum cases, the monopole motion period increases with temperature (or entropy)
and that for a given temperature or entropy, this time increases with the
initial compression. The situation is not as clear in the semi-classical
case mainly due to the presence of large fluctuations in the central density
which makes the extraction of the corresponding times difficult.

Therefore, as far as the collective motion is concerned, we can conclude
that, compared to the semi-classical predictions, the quantum simulations
present a faster expansion towards lower densities. These differences
appear to be due to the differences in the collective potential but also to
the differences in the damping and cooling processes which appear faster in
the semi-classical case. Then, in order to conclude this comparison let us
discuss in more details the evaporation process.

\subsubsection{Evaporation dynamics and time scales}

\smallskip

\begin{figure}[tbph]
\begin{center}
\includegraphics*[height=16cm,width=16cm]{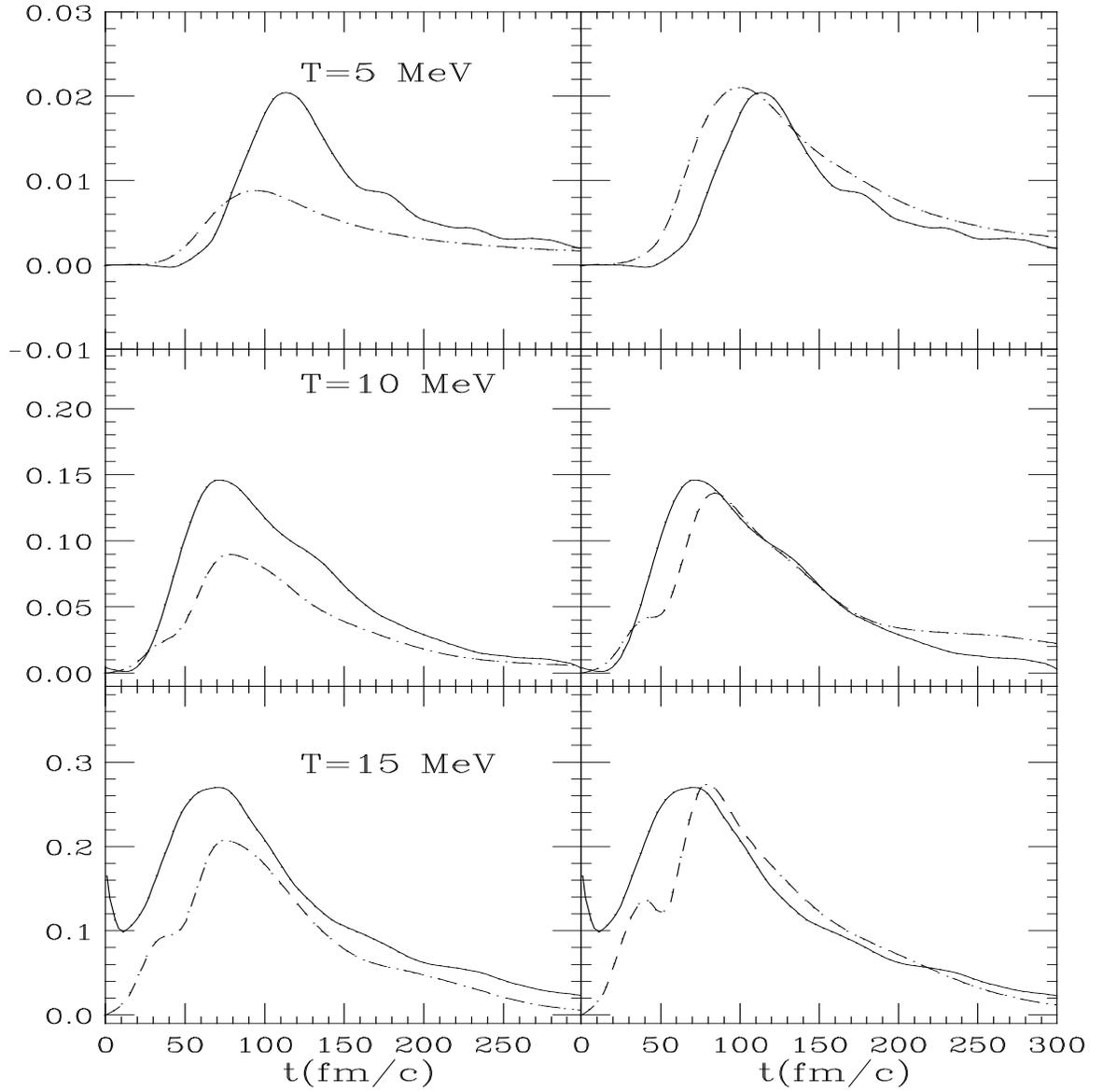}
\end{center}
\caption{Particle Flow out of a sphere of radius 15 fm for a typical Ca
nucleus at various temperatures $T=5,$ $10$ and $15$ MeV (from top to
bottom). Left: Quantum without surface (dashed line). Right: Quantum+Surface
(dashed line). On each figures is superimposed the Semi-classical case
(solid line).}
\label{fig:9}
\end{figure}

In order to study the evaporation process we have computed the particle flows
outside of a sphere of radius 15 fm. These flow are shown on figure (\ref
{fig:9}). 

\begin{figure}[tbph]
\begin{center}
\includegraphics*[height=10cm,width=10cm]{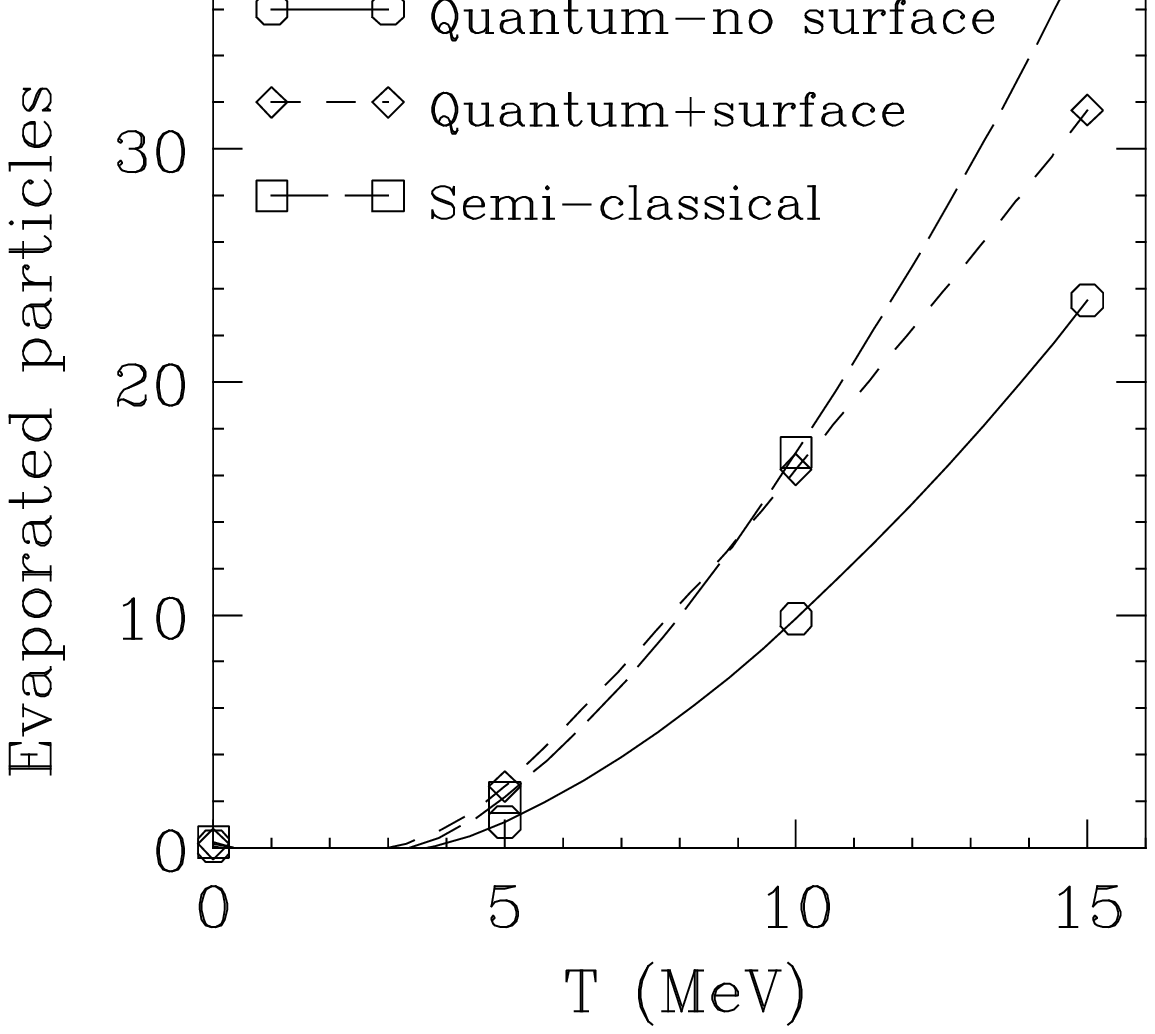}
\end{center}
\caption{Asymptotic number of evaporated particles (after $300fm/c$) as a
function of the temperature for the different considered models. Solid line:
Quantum without surface. Dashed line: Quantum+Surface. Long-dashed line:
Semi-classical.}
\label{fig:evap2}
\end{figure}

According to this figure it is clear that the semi-classical approximation
predicts a quicker evaporation of more particles than the quantum one. We
see that these differences in the evaporation process are reduced by the
introduction in the quantum simulation of a surface interaction (see right
part of Fig.\ref{fig:9}). 
This can be clearly quantified by the asymptotic number of evaporated
particles (after $300fm/c$) which is presented in Fig. \ref{fig:evap2} as a
function of the temperature/entropies for different simulations. We can see
that the temperature acts completely differently in quantum and
semi-classical cases. In particular for high temperature, the semi-classical
model is poor approximation of the quantum mean-field theory. The
heated nucleus appears less bound in semi-classical than in quantum cases.
However, we note that the introduction of a surface term reduces
quantitatively the gap between the two treatments. Nevertheless, the time
scales of the evaporation are different.

This important difference can be traced back to the fact that in a
quantum approach the nucleons can be partly reflected by the mean-field
potential wells while in a classical case as soon as a particle as enough
energy it will be immediately emitted. This large difference between
semi-classical and quantum approaches is one of the major drawback of
semi-classical approximation. Indeed, it affects not only the evaporation
itself but in fact the whole dynamics changing time-scales, cooling
processes, influence of the thermal pressure and also the one-body
dissipation.

\section{Discussion of the expansion and spinodal instabilities}

\smallskip


Semi-classical simulations (including fluctuations) of the spinodal
decomposition are able to reproduce globally the experimental fragment
partitions\cite{Sur92,Gua1}. However, these simulations seem to underestimate the
kinetic energies of the fragments. This is a general observation made on
BUU simulations that the expansion velocity at the entrance of the spinodal
region is barely enough compared with the experimental data leaving
small room for additional slowing down during the fragment formation. It is
thus important to discuss the differences observed between
semi-classical and quantum dynamics when the nuclei reach low density
regions.

\smallskip


At zero temperature, we have seen that the infinite system becomes unstable
when the density is less than $0.1fm^{-3}$. 

\begin{figure}[tbph]
\begin{center}
\includegraphics*[height=10cm,width=10cm]{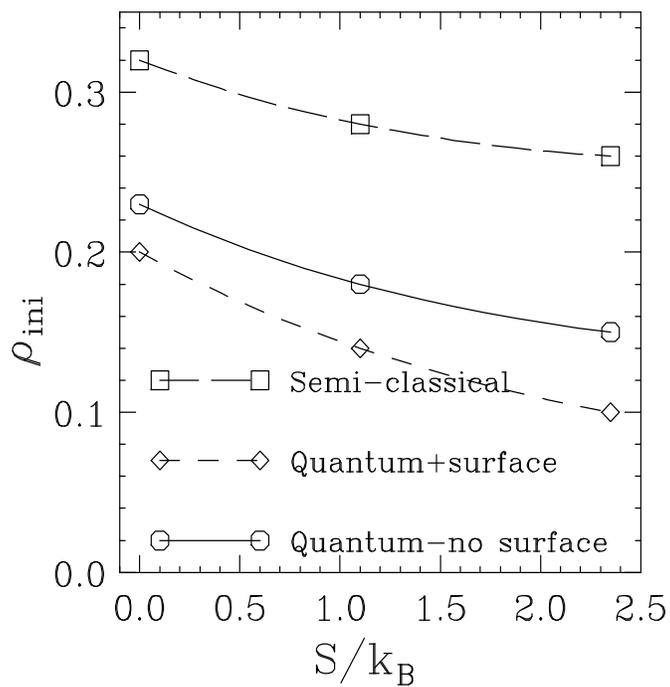}
\end{center}
\caption{Initial density that gives a minimal density equal to $0.1$ fm$^{-3}
$ as a function of the entropy. Solid-line: Quantum (without surface).
Dashed-line: Quantum+Surface. Long dashed line: Semi-classical.}
\label{fig:rho01}
\end{figure}

\begin{figure}[tbph]
\begin{center}
\includegraphics*[height=10cm,width=10cm]{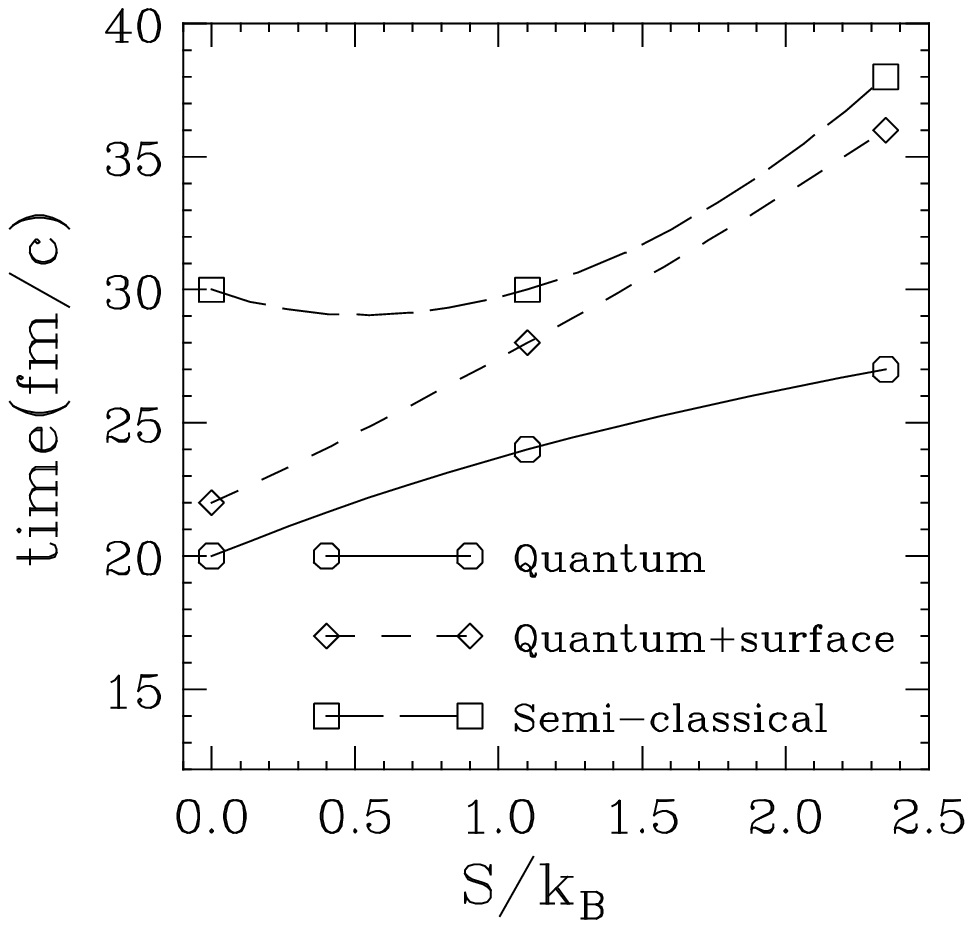}
\end{center}
\caption{Time necessary to reach a minimal density equal to $0.1$ fm$^{-3}$
as a function of the entropy (for the particular initial density that has
the minimal density at $0.1$ fm$^{-3}$. Solid-line: Quantum (without
surface). Dashed-line: Quantum+Surface. Long dashed line: Semi-classical.}
\label{fig:time01}
\end{figure}
From figure \ref{fig:rho01}, 
it is clear that higher densities (up to a factor 2)
are needed in semi-classical calculations to reach a density below $0.1$ $%
fm^{-3}$. If we look now at figure \ref{fig:time01}, we see that the time
needed to reach the low densities is also longer in semi-classical
simulation. This should be related to the observed stronger damping of
collective motion in the Vlasov case. The induced delay of a factor 1.5 to 2
is associated with a reduction of the collective expansion velocity by the
same amount. This may explain why semi-classical simulations taking into
account the fragment formation dynamics seem to underestimate the
experimental kinetic energies. From this study, it is now clear that quantal
simulations are important when one wants to discuss quantitatively
collective expansions
and fragment formation that could occur during Heavy Ion reactions.


\section{Concluding remarks}

\smallskip

This work presents a comparison between quantum and semi-classical
mean-field dynamics of hot and compressed nuclei.

\smallskip We have first pointed out some properties of the semi-classical
approaches using test particle sampling of the phase space. One
particularity of this method is the smoothing or the coarse-graining of
the phase space in order to define the density. 
This smoothing adds a
finite range term to the initial potential (expression (\ref{Eq:10})). This
numerical surface term affects
static and dynamical properties of
nuclei. We have disentangled effects due to this extra surface term from
intrinsic differences due to the absence of quantum features in
semi-classical simulation by comparing them with quantum calculations in
which a short range potential was added in order to mimic the
numerical one present in semi-classical approaches.

\smallskip From these three different types of simulations (semi-classical,
quantum with and without finite range interaction) we have drawn the
conclusions listed bellow.

\begin{itemize}
\item  \smallskip While the three approaches have identical static
properties of infinite nuclear matter, the energies of finite size nuclei at
different dilutions and excitations are different. In particular,
the compressibility modulus (or equivalently the curvature of the 
EOS) of the finite systems depends upon
the considered approach. This directly affects the collective dynamics in
nuclei. In particular, we have shown that the frequencies of the breathing
mode, the dependence in temperature of the amplitude of the collective modes
and the position of stationary points are affected by the presence
of a numerical gradient of density in the potential. However, the presence
of this numerical surface term do not explain all the observed differences
and part of the discrepancy is also due to the semi-classical approximation.

\item  \smallskip From the dynamical evolutions, 
we can also see that the damping
of the collective mode is greater in semi-classical method than in quantum
approaches despite the absence of Landau spreading of the breathing mode 
\cite{mono} in semi-classical approaches. This strong damping observed in
semi-classical simulations reflects the differences in the evaporation
process as well as additional dissipation in the classical motion of
particles. In the semi-classical case, highly excited particles are emitted
early in the evolution inducing a large energy loss during the initial
stage of the dynamics. This implies a fast attenuation of the collective
mode. Conversely, in the quantum case, the reflections of the
single-particle wave functions on the potential well reduce the particle
evaporation rate and increase the pressure on the nucleus boundary. This
difference on time scale is smaller but remains important when a surface
interaction is introduced in the quantum-treatment. The evaporation is
faster and the total number of emitted particles is larger in semi-classical
treatments compared with quantum approaches.

\item  \smallskip The differences in the evaporation process not only affect
the damping of the monopole vibration but they also change the size of the
remaining residue after desexcitation. At low temperature but extreme
conditions of compression, the fast evaporation observed in semi-classical
approaches gives a strong loss of energy that enable the system to survive.
On the contrary, at high energy, in the semi-classical case, the potential
is not able to retain the particles inside the nucleus: the system is
directly vaporized. In the quantum case, at low temperature, the collective
motion is less damped so that the expansion leads more easily to a complete
dilution of the system. Conversely, at high temperature, the evaporation is
reduced by the barrier reflections so that a non compressed nucleus may
survive the evaporation stage. In summary, in quantum approaches the mean-field is
more robust.

\item  \smallskip Finally, we have discussed that in a semi-classical
treatment the expansion is slower and that stronger compression are needed
in order to reach low density regions. This induces an underestimation
of the expansion velocity by a factor 1.5 to 2. This may fill the gap
between theoretical prediction of fragment kinetic energies and experimental
data since up to now stochastic mean-field description of multifragmentation
have been always performed using a semi-classical approximation.
\end{itemize}

{\bf Acknowledgements}

We thank D. Durand for helpful discussion about the physical content of
the paper.

\end{document}